\documentclass[twocolumn]{aastex631}

\newcommand{\Mjup}{\mbox{$\rm M_\mathrm{Jup}$}}

\newcommand{\Rsun}{\mbox{$\rm R_{\odot}$}}

\newcommand{\e}[1]{\times 10^{#1}}

\shorttitle{51~Eri is a Gamma~Doradus Pulsator}
\shortauthors{Sepulveda et al.}

\begin{document}
\title{The Directly-Imaged Exoplanet Host Star 51~Eridani is a Gamma~Doradus Pulsator}

\correspondingauthor{Aldo G. Sepulveda}
\email{aldo.sepulveda@hawaii.edu}
\author[0000-0002-8621-2682]{Aldo G. Sepulveda}
\altaffiliation{NSF Graduate Research Fellow}
\affiliation{Institute for Astronomy, University of Hawai`i at M\={a}noa. 2680 Woodlawn Drive, Honolulu, HI 96822, USA}

\author[0000-0001-8832-4488]{Daniel Huber}
\affiliation{Institute for Astronomy, University of Hawai`i at M\={a}noa. 2680 Woodlawn Drive, Honolulu, HI 96822, USA}

\author[0000-0002-3726-4881]{Zhoujian Zhang} 

\affiliation{Department of Astronomy, The University of Texas at Austin,
2515 Speedway Blvd. Stop C1400 
Austin, TX 78712, USA}
\affiliation{Institute for Astronomy, University of Hawai`i at M\={a}noa. 2680 Woodlawn Drive, Honolulu, HI 96822, USA}

\author[0000-0001-9313-251X]{Gang Li}
\affiliation{IRAP, Université de Toulouse, CNRS, CNES, UPS, Toulouse, France}

\author[0000-0003-2232-7664]{Michael C. Liu}
\affiliation{Institute for Astronomy, University of Hawai`i at M\={a}noa. 2680 Woodlawn Drive, Honolulu, HI 96822, USA}

\author[0000-0001-5222-4661]{Timothy R.\ Bedding} %
\affiliation{Sydney Institute for Astronomy (SIfA), School of Physics, University of Sydney, NSW 2006, Australia}
\affiliation{Stellar Astrophysics Centre (SAC), Department of Physics and Astronomy, Aarhus University, Ny Munkegade 120, DK-8000 Aarhus C, Denmark}

\submitjournal{The Astrophysical Journal}
\accepted{September 12, 2022}

\begin{abstract}
51~Eri is well known for hosting a directly-imaged giant planet and for its membership to the $\beta$~Pictoris moving group. Using two-minute cadence photometry from the \textit{Transiting Exoplanet Survey Satellite} (\textit{TESS}), we detect multi-periodic variability in 51~Eri that is consistent with pulsations of Gamma~Doradus ($\gamma$~Dor) stars. We identify the most significant pulsation modes (with frequencies between $\sim$0.5--3.9 cycles/day and amplitudes ranging between $\sim$1--2 mmag) as dipole and quadrupole gravity-modes, as well as Rossby modes, as previously observed in \textit{Kepler} $\gamma$~Dor stars. Our results demonstrate that previously reported variability attributed to stellar rotation is instead likely due to $\gamma$~Dor pulsations. Using the mean frequency of the $\ell =1$ gravity-modes, together with empirical trends of the \textit{Kepler} $\gamma$~Dor population, we estimate a plausible stellar core rotation period of 0.9$^{+0.3}_{-0.1}$~days for 51~Eri. We find no significant evidence for transiting companions around 51~Eri in the residual light curve. The detection of $\gamma$~Dor pulsations presented here, together with follow-up observations and modeling, may enable the determination of an asteroseismic age for this benchmark system. Future \textit{TESS} observations would allow a constraint on the stellar core rotation rate, which in turn traces the surface rotation rate, and thus would help clarify whether or not the stellar equatorial plane and orbit of 51~Eri~b are coplanar. 
\end{abstract}


\section{Introduction}
51~Eridani (51~Eri, HIP~21547, HD~29391) is a F0~IV star \citep{Abt+Morrell1995} with a $V$-band magntiude of 5.2 \citep{Hog+2000} located at a distance of 29.91$\pm$0.07 pc \citep{gaiamission,GaiaEDR3}. It hosts a directly-imaged giant planet \citep[51~Eri~b;][]{Macintosh+2015,DeRosa+2015} with a semi-major axis of $\sim$10--13~au, a moderate eccentricity of $\sim$0.5, and a mass of $\lesssim$11~\Mjup \citep[e.g.,][]{Maire+2019,DeRosa+2020,Bowler+2020,Dupuy+2022}. In addition to hosting the imaged giant planet, 51~Eri  is a member of the $\beta$~Pictoris moving group \citep[$\beta$PMG;][]{Zuckerman+2001,Malo+2013}, which implies an age of $\sim$19--24 Myr for the star \citep{Bell+2015,MiretRoig+2020}. Another member of the $\beta$PMG, GJ~3305~AB, is an M~dwarf binary gravitationally bound to 51~Eri at projected separation of $\sim$1990~au \citep{Feigelson+2006,Kasper+2007}. Furthermore, 51~Eri has an infrared excess that is consistent with a cold debris disk \citep{RiviereMarichalar+2014}, but resolved imaging of this disk has thus far remained elusive due to its intrinsic faintness \citep[e.g.,][]{Pawellek+2021}. 

A significant challenge is the uncertain system age, which affects mass estimates of 51~Eri~b based on substellar cooling models \citep[e.g.,][]{Samland+2017,Rajan+2017}. As dynamical mass measurements of imaged substellar companions increase thanks to \textit{Gaia} astrometry \citep[e.g.,][]{Brandt+2021,Dupuy+2022,Franson+2022}, precise ages are then needed to compare these dynamical masses to masses predicted from different cooling models and thereby test hot/warm/cold-start formation scenarios \citep[e.g.,][]{Marley+2007,Spiegel+Burrows2012,Mordasini2013}. Age estimates of the $\beta$PMG range from $\sim$8--40~Myr \citep[e.g.,][ their Table~1]{Mamajek+Bell2014}. Additionally, age estimates for 51~Eri itself include both younger and older ages than the current nominal $\beta$PMG age of $\sim$19--24~Myr. For example, \citet{Simon+Schaefer2011} measured the angular diameter of 51~Eri and combined with stellar evolution models estimated an age of 13$\pm$2~Myr. \citet{Montet+2015} combined their dynamical mass measurement of GJ~3305~AB with evolution models to derive an age of 37$\pm$9~Myr, which also applies to 51~Eri assuming the three stars formed at the same time. Precise, independent age estimates are key to further quantifying the properties of this system as well as determining which substellar cooling models are most consistent with 51~Eri~b.

Asteroseismology probes stellar interiors and is a powerful tool to determine ages \citep[e.g., review by][]{kurtz22}. Early F-type stars show gravity-mode pulsations with periods between 0.3 to 3~days, forming the class of $\gamma$\,Doradus variables \citep{Balona+1994,Kaye+1999}. The $\gamma$\,Dor stars are located between the $\delta$\,Scuti stars, which lie in the classical instability strip, and solar-like oscillators, with some hybrid $\delta$\,Scuti and $\gamma$\,Dor stars pulsating in pressure and gravity-modes \citep{Grigahcene+2010}. The \textit{Kepler} Mission \citep{Borucki+2010} led to the discovery that nearly all $\gamma$~Dor stars pulsate in dipole modes and enabled the measurement of their core rotation rates \citep{vanreeth16,li20}. $\gamma$\,Dor pulsations have been detected in the directly-imaged exoplanet host star HR\,8799 \citep{Marois+2008,Marois+2010} using ground-based observations \citep{zerbi99} and were used to constrain an asteroseismic age \citep[][albeit with an ambiguity related to uncertainty in the stellar inclination]{moya10}. However, subsequent space-based data from the \textit{Microvariability and Oscillations of STars} (\textit{MOST}) telescope and from the \textit{BRIght Target Explorer} (\textit{BRITE}-Constellation) questioned the mode identification and implied only a single independent frequency, limiting the potential for asteroseismology in HR~8799 \citep{sodor14,Sodor+Bognar2020}. The unambiguous detection of multi-periodic pulsations in 51~Eri would open the door for determining an asteroseismic age for this system.

\citet{Koen+Eyer2002} searched for candidate photometric variables using \textit{Hipparcos} $V$-band observations and reported 51~Eri as a probable ``microvariable." \citet{Desidera+2021} recently investigated Sector 5 \textit{Transiting Exoplanet Survey Satellite} (\textit{TESS}, \citealt{TESSMission}) photometry of 51~Eri and noted multiple periodicities likely due to stellar pulsations, but provided no further interpretation of 51~Eri's variability status. Here we present Sector 5 and Sector 32 \textit{TESS} photometry of 51~Eri and demonstrate that it is a $\gamma$~Dor variable. We extract significant frequencies from the light curve in Section \ref{sec:obsAndFreq} and interpret the results in Section \ref{sec:discussion}. We summarize our findings and recommendations in Section \ref{Sec:Conclusion}.

\section{\textit{TESS} Observations and Frequency Extraction}\label{sec:obsAndFreq}
\begin{figure*}[h!]
  \includegraphics[trim=0cm 0cm 0cm 0cm, clip, width=7.09in]{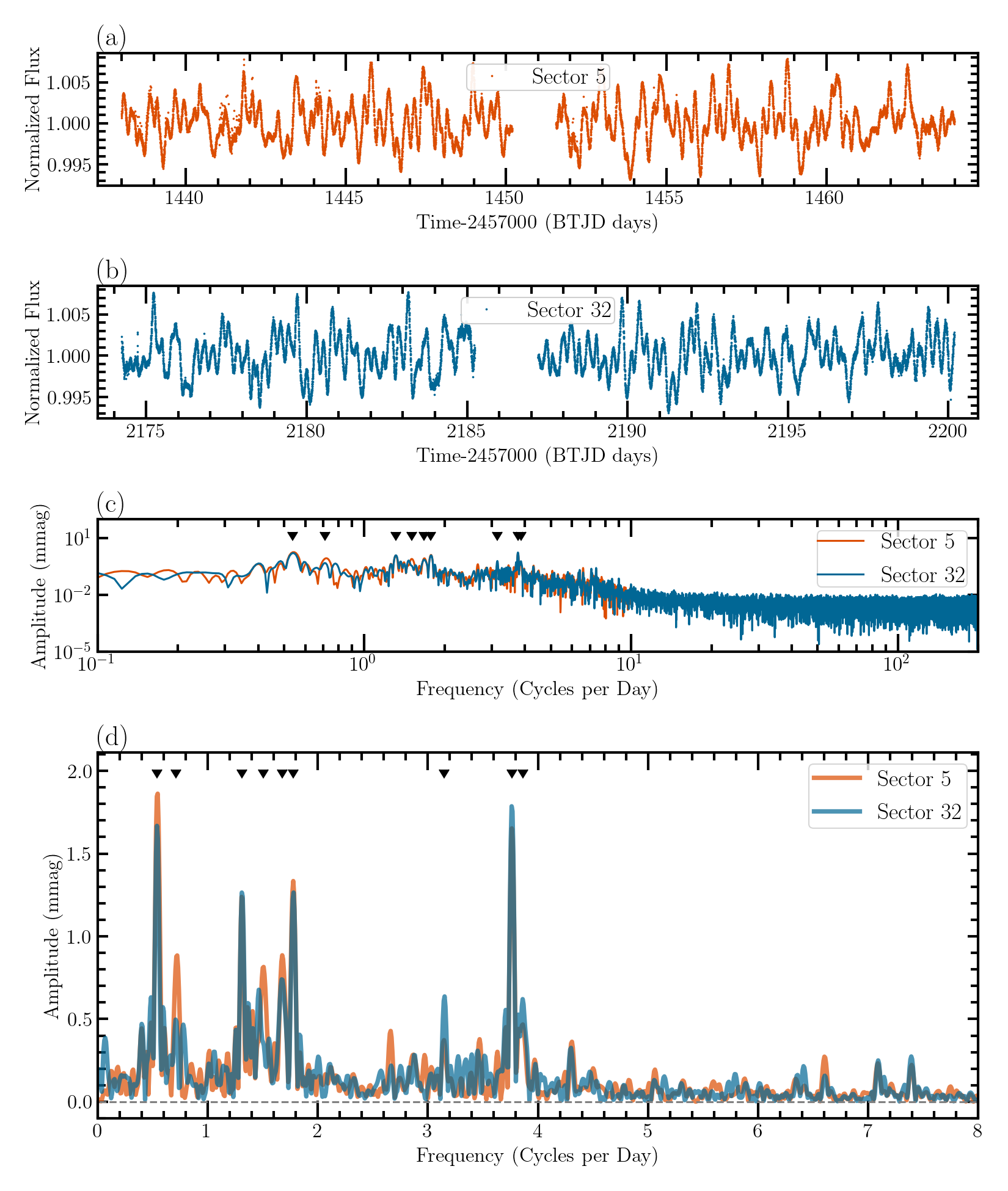}
  \centering
  \caption{(a): \textit{TESS} Sector 5 light curve of 51~Eri. Error bars are smaller than the symbol sizes. (b): Same as (a) but for Sector 32. (c): Corresponding amplitude spectra of each light curve in log-log space. (d): Corresponding amplitude spectra of each light curve in linear-linear space and zoomed in on the location of the $\gamma$~Dor pulsations. Downward triangles denote the significant pulsation frequencies that we extracted from the concatenated time series as described in Section \ref{subsec:freqExtract}.
  \label{fig:51EriTESS}} 
\end{figure*}

We downloaded all available \textit{TESS} two-minute cadence observations of 51~Eri (Sectors 5 and 32) using the \texttt{lightkurve} package \citep{LightkurveCollaboration+2018}. We used the PDC-SAP light curves \citep{smith12,stumpe12,stumpe14} provided by the Science Processing Operations Center \citep[SPOC,][]{jenkins16} and removed outlying photometry using the built-in routines of \texttt{lightkurve}, resulting in 34,908~cadences. The light curves from both sectors show clear evidence for multi-periodic variability (Figure \ref{fig:51EriTESS}a and \ref{fig:51EriTESS}b). We calculated amplitude spectra of each sector, which show consistent pulsation frequencies between $\sim$0-8~cycles/day (Figure \ref{fig:51EriTESS}c and Figure \ref{fig:51EriTESS}d). The amplitude differences between the two Sectors (on the order of tenths of mmag) are not surprising given that the observed peaks are expected to comprise individual radial orders \citep[e.g.,][]{li20} that are not fully resolved with the current \textit{TESS} data.

\begin{figure*}[t!]
  \includegraphics[trim=0cm 0cm 0cm 0cm, clip,width=7.09in]{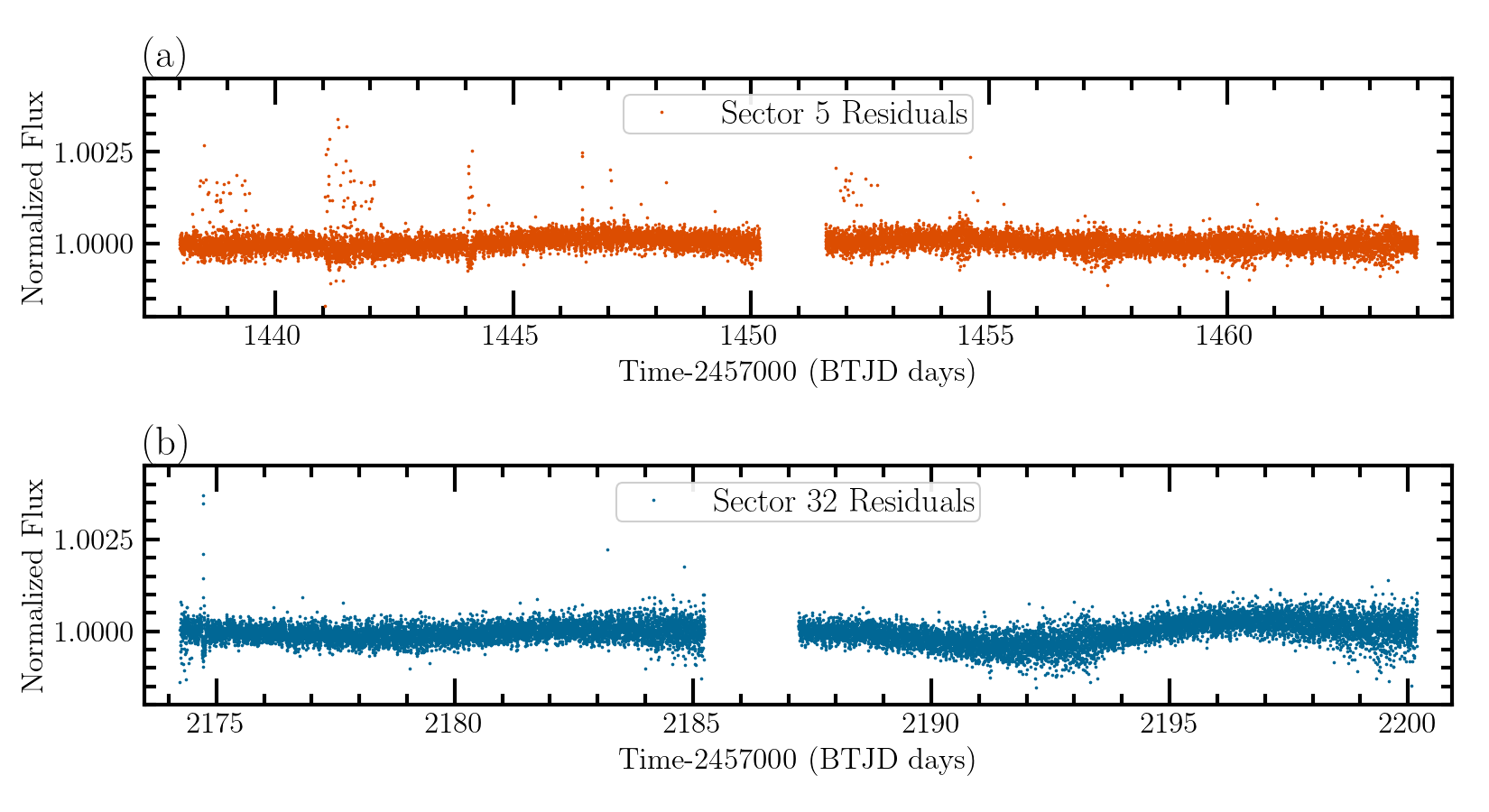}
  \centering
  \caption{Residual \textit{TESS} light curve of 51~Eri after removing high-frequency signal as described in Section \ref{subsec:residuals}. No significant non-stellar variability is detected. \label{fig:residuals}} 
\end{figure*}
\subsection{Pulsation Frequency Extraction}\label{subsec:freqExtract}
We extracted significant pulsation frequencies using the \texttt{SigSpec} package \citep{SigSpec}, which iteratively fits sine-waves to the time series up to our defined spectral significance threshold. To inform our threshold choice, we check the data for red noise, which describes a noise structure that increases towards lower frequencies and is typically a consequence of stellar granulation or instrumental effects. The amplitude spectra in log-log space (Figure \ref{fig:51EriTESS}c) show red noise at $\lesssim$12~cycles/day. We interpret this as evidence for granulation due to a thin surface convection zone on 51~Eri, as found on some intermediate mass $\delta$\,Scuti pulsators \citep[e.g.,][]{Kallinger+Matthews2010}. Alternatively, the red noise may also be due to instrumental effects. To mitigate the effect of the red noise, we chose a stringent spectral significance threshold of 500 (corresponding to a signal to noise ratio of $\sim$40) for the purpose of identifying significant pulsation modes.

We performed the frequency extraction on the concatenated time series using \texttt{SigSpec} from 0.2--24 cycles/day. This yielded nine significant frequencies that meet our spectral significance threshold and they are displayed in Figures \ref{fig:51EriTESS}c and \ref{fig:51EriTESS}d. A list of these significant frequencies and their corresponding amplitudes are provided as Table \ref{tab:freqs}. Phase angles, which serve as a zero-point for the sine waves, are also included for completeness.\footnote{While not used in this study, phase angles are necessary for constructing model light curves from the detected pulsation frequencies and they are also useful quantities when performing mode identification with multi-color photometry.} Uncertainties were calculated following \citet{Kallinger+2008}. Performing the frequency extraction on each Sector separately yields similar results. 

Figures \ref{fig:51EriTESS}c and \ref{fig:51EriTESS}d show that there are likely more pulsation frequencies in these data than those reported in Table \ref{tab:freqs}. A more careful treatment of the red noise would be required to better assess their significance. In particular, Figures \ref{fig:51EriTESS}c and \ref{fig:51EriTESS}d show evidence for signal between 6--8~cycles/day, which may be pressure-modes or combination frequencies of gravity-modes. In the pressure-modes scenario, this would indicate that 51\,Eri is a hybrid $\gamma$\,Dor-$\delta$\,Scuti pulsator.

\begin{deluxetable*}{ccccc}[t!]
\tablehead{
\colhead{Frequency} & \colhead{Amplitude} & \colhead{Phase Angle} & \colhead{Spectral Significance\tablenotemark{a}}&\colhead{Preliminary Mode-ID}\\
\colhead{(cycles/day)} & \colhead{(mmag)} &\colhead{(rads)}&\colhead{}& \colhead{}\\ }
\caption{Significant pulsation frequencies of 51~Eri. \label{tab:freqs}} 
\startdata
0.5398$\pm$0.0005&1.77$\pm$0.04&1.99$\pm$0.01&1621&Rossby mode\\
3.7659$\pm$0.0004&1.66$\pm$0.04&4.65$\pm$0.01&1621&$\ell =2$ g-mode\\
1.7764$\pm$0.0005&1.33$\pm$0.03&1.44$\pm$0.01&1511&$\ell =1$ g-mode\\ 
1.3094$\pm$0.0005&1.25$\pm$0.03&0.39$\pm$0.01&1511&$\ell =1$ g-mode\\ 
1.6762$\pm$0.0006&0.80$\pm$0.02&2.58$\pm$0.01&1222&$\ell =1$ g-mode\\
0.7132$\pm$0.0006&0.90$\pm$0.03&2.55$\pm$0.02&958&Rossby mode\\
1.5052$\pm$0.0007&0.75$\pm$0.03&5.87$\pm$0.02&723&$\ell =1$ g-mode\\
3.8616$\pm$0.0007&0.55$\pm$0.02&0.70$\pm$0.02&674&$\ell =2$ g-mode\\
3.1491$\pm$0.0008&0.90$\pm$0.04&6.05$\pm$0.02&600&$\ell =2$ g-mode\\
\enddata
\tablenotetext{a}{Formally we report the cumulative spectral significance for this column. For a more detailed description, see \citet{Reegen+2011}.}
\end{deluxetable*} 
\subsection{Probing for Non-Stellar Variability}\label{subsec:residuals}
We repeated the frequency extraction procedure in an identical manner but using a spectral significance threshold of 10 for the purpose of removing signal that is intrinsically associated with the star. We inspected the corresponding residual light curve (Figure \ref{fig:residuals}) for any clear indications of brightness ``dips" that may be due to transiting companions or infalling bodies around 51~Eri \citep[e.g.,][]{Zieba+2019,Hey+2021}. Such features could plausibly be obscured by the $\gamma$~Dor variability and red noise in the original light curve and thus may be evident after a more aggressive frequency subtraction. Given our lower frequency extraction limit of 0.2 cycles/day, these residuals thereby probe for transiting companions with a period longer than 5 days.

No unambiguous brightness dips are evident in this residual light curve which corresponds to a nondetection of transiting companions in these data. We evaluated this using Box Least Squares \citep[BLS,][]{Kovacs+Zucker2002} periodograms where we found no significant signal consistent with a transit. Instead, we note a few brightness excess features that may either be systematic artifacts or indicative of stellar flares (e.g., at $\sim$1446.45 and $\sim$2174.73 days). We also note a trend of long-period variation on the order of $\sim$14 days that likely corresponds to an instrumental artifact related to the orbital period of the \textit{TESS} satellite.

In the scenario where the brightness excess features are flares, they would more likely originate from a contaminating source than from 51~Eri itself. Early F-type stars like 51~Eri are generally not expected to be active due in part to a lack of deep convective envelope \citep[e.g.,][]{Charbonneau2010,Brun+Browning2017}. There is some contention in the literature due to a small but growing number of candidate flaring A and F stars that challenge this paradigm \citep[e.g.,][]{Balona2012,Balona2015} while, on the other hand, such cases are sometimes explained by contamination from a source other than the respective star \citep[e.g.,][]{Pedersen+2017}. \citet{Antoci+2019} noted a flare in the \textit{TESS} data for the $\gamma$~Dor star $\pi$~PsA, although they suggest it likely originates from a background star or a bound companion and not from $\pi$~PsA itself. While a more thorough investigation for 51~Eri is beyond the scope of this study, we note that the bound M-Dwarf binary GJ~3305 \citep[with a $V$-band magnitude of 10.6,][]{Reid+2004} is separated from 51~Eri by 66~\arcsec \citep{Feigelson+2006}. This makes GJ~3305 a likely source of contamination in the \textit{TESS} aperture that we used for 51~Eri and thus a likely candidate to explain any observed flares.

\section{Discussion}\label{sec:discussion}
\subsection{$\gamma$ Dor Classification and Preliminary Mode Identification}\label{subsec:classification}
Our nine extracted pulsation frequencies from the \textit{TESS} photometry (Table \ref{tab:freqs}) are consistent with those of $\gamma$\,Dor pulsators. Following the patterns identified for $\sim$600 \textit{Kepler} $\gamma$\,Dor stars from \citet{li20}, we interpret the peaks near 0.26\,d (3.8~cycles/day) as quadrupole ($\ell =2$) gravity-modes and the peaks at 0.5--0.8\,d (1.3--1.8~cycles/day) as dipole ($\ell =1$) gravity-modes. Typical gravity-mode period spacings, which enable identification of radial orders and constrain core rotation rates, are on the order of hundreds of seconds and thus cannot be reliably resolved with only two sectors of \textit{TESS} data. The longest-period peak near 1.9\,d (0.54~cycles/day) may be associated with Rossby modes, which are observed in $\sim$\,13\% of all \textit{Kepler} $\gamma$\,Dor stars \citep{li20}. This preliminary mode identification is included in Table \ref{tab:freqs}.

\begin{figure}[th!]
  \includegraphics[trim=0cm 0cm 1.25cm 1.25cm, clip,width=3.35in]{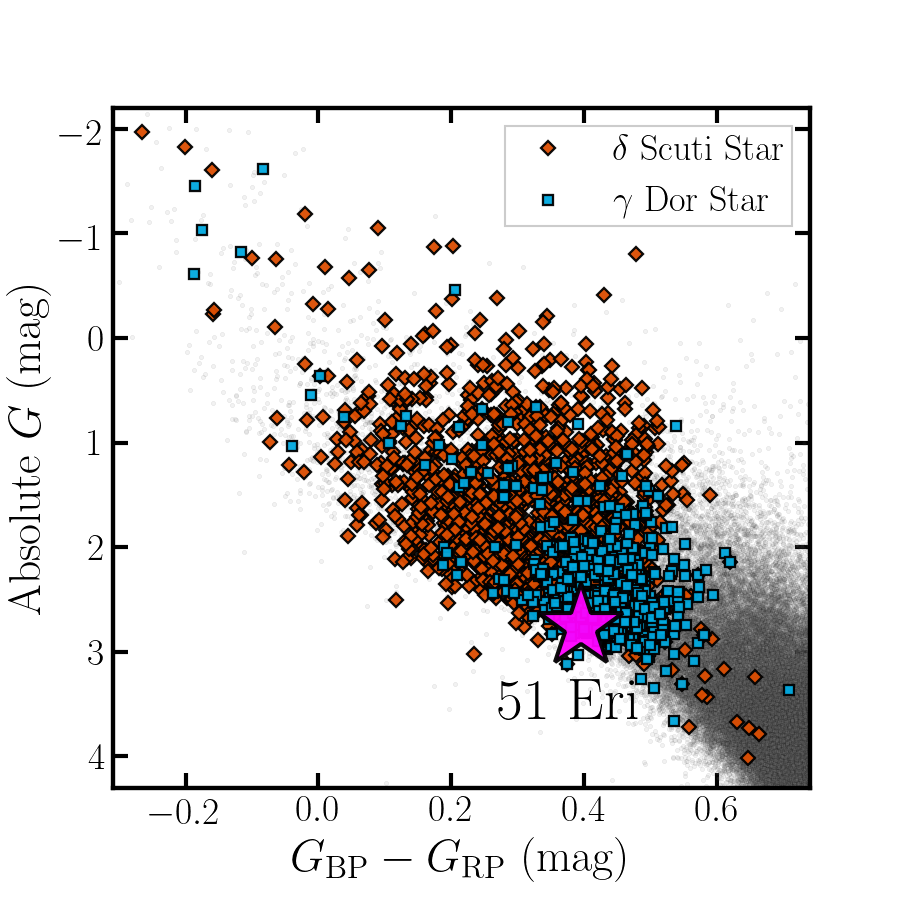}
  \centering
  \caption{\textit{Gaia}-based color-magnitude diagram for a sample of \textit{Kepler} stars as described in Section \ref{subsec:classification}. The orange diamonds are the $\delta$~Scuti stars and the blue squares are the $\gamma$~Dor stars overplotted. 51~Eri is overplotted as a large magenta star. The small, grey, semi-transparent dots represent the general population of \textit{Kepler} stars.  \label{fig:colormag}} 
\end{figure}

To further illustrate the $\gamma$ Dor classification of 51~Eri, we constructed a color-magnitude diagram to compare it to other known $\gamma$~Dor and $\delta$~Scuti pulsators from the \textit{Kepler} mission. The sample is based on the \textit{Kepler} Input Catalog (KIC, \citealt{Brown+2011}) cross-matched with \textit{Gaia} Early Data Release 3 \citep[EDR3,][]{gaiamission,GaiaEDR3} and we calculated the absolute $G$ magnitude using inverse \textit{Gaia} EDR3 parallax measurements as the distances. To account for dust reddening and extinction, we utilized the $V$-band extinctions derived from 3D dust maps for the KIC stars \citep{Green+2019,Berger+2020} and we converted their values to the corresponding \textit{Gaia} passbands using the relative extinction ratios from \citet{Wang+Chen2019}. We excluded any stars with a \textit{Gaia} EDR3 parallax uncertainty of $>20\%$. The subsample of \textit{Kepler} stars that we denote as $\delta$~Scuti pulsators are from \citet{Murphy+2019} and the subsample that we denote as $\gamma$~Dor pulsators are from \citet{li20}. Extinction is negligible for 51~Eri \citep[e.g.,][]{Guarinos1992}, and its $BP-RP$ color and absolute $G$ magnitude are indeed consistent with these other $\gamma$~Dor and $\delta$~Scuti stars (Figure \ref{fig:colormag}), further supporting its $\gamma$~Dor classification.

\subsection{On the Rotation Period of 51~Eri}
\citet{Koen+Eyer2002} derived a peak frequency of 1.5365~cycles/day using \textit{Hipparcos} photometry, which has nominally been associated with a stellar rotation period of 0.65~days. \citet{Maire+2019} and \citet{Desidera+2021} recovered the same period from the \textit{Hipparcos} data and with ground-based photometry from the Multi-site All-Sky CAmeRA (MASCARA). Assuming the variability is associated with stellar rotation, \citet{Maire+2019} combined it with $v\sin{i}$ and stellar radius estimates to infer a spin-axis inclination of $\sim$41--45$^{\circ}$, which is similar to the estimate of $\sim$45$^{\circ}$ reported by \citet{Feigelson+2006} using the same period and methods. Notably, these stellar spin-axis inclination estimates based on the 0.65 day rotation period are similar to the orbital inclination of 51~Eri~b within its uncertainties \citep[e.g.,][]{Maire+2019,DeRosa+2020,Bowler+2020,Dupuy+2022}. 

The \textit{TESS} data shows a significant peak at 1.5052$\pm$0.0007 cycles/day (Figure \ref{fig:51EriTESS}d) that recovers this previously reported \textit{Hipparcos}/MASCARA frequency within the $1\sigma$ error bars reported by \citet{Maire+2019}. However, in our analysis this frequency is not among the six most significant frequencies (Table \ref{tab:freqs}). Moreover, the amplitude from \textit{TESS} for this frequency is only 0.75$\pm$0.03 mmag, whereas \citet{Koen+Eyer2002} report a \textit{Hipparcos V}-band amplitude of 5.3\,mmag. A likely explanation is that the \textit{Hipparcos} and MASCARA frequency was comprised of unresolved $\ell =1$ dipole gravity-modes that required the continuous 2-minute cadence of \textit{TESS} to resolve. We conclude that this frequency is most likely due to  gravity-mode pulsations, which means the stellar rotation period of 51~Eri is presently undetermined. Because this nominal rotation period was previously used to estimate the stellar spin-axis inclination \citep[e.g.,][]{Feigelson+2006,Maire+2019}, the coplanarity of the stellar equatorial plane with the orbit of 51~Eri~b is presently unclear.\footnote{Consistent line-of-sight inclinations are a necessary but insufficient criteria for two planes to be mutually coplanar. Therefore, significantly differing inclinations can serve to rule out coplanarity, but similar inclinations on their own are consistent with, but do not confirm, coplanarity.}

\begin{figure*}[t]
  \includegraphics[trim=3.85cm 0cm 3.85cm 1.5cm, clip,width=7.1in]{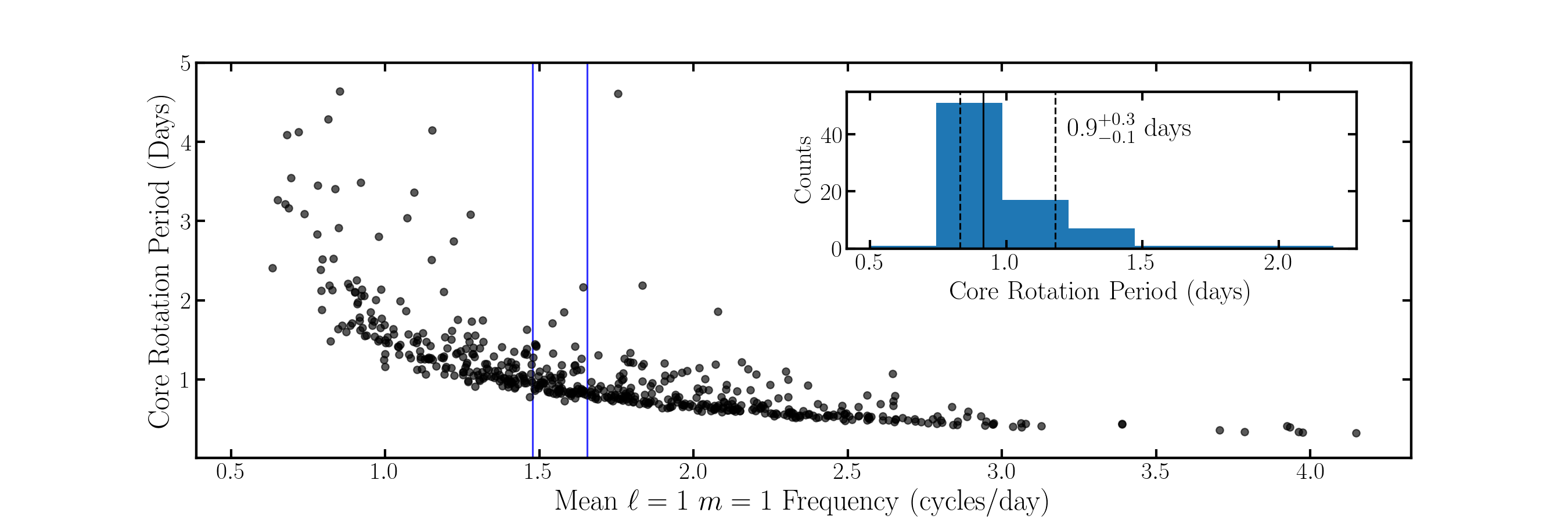}
  \centering
  \caption{\textit{Kepler} $\gamma$~Dor stars from \citet{li20} that pulsate in $\ell =1$ gravity-modes and that also have a core rotation measurement. The black dots denote their mean $\ell =1$ gravity-mode frequency compared to their core rotation period. The vertical blue lines denote the boundaries of our estimate for the mean $\ell =1$ gravity-mode frequency of 51~Eri (1.57$\pm$0.09 cycles/day) which is used to collect a subpopulation of core rotation periods. The inset histogram corresponds to this subpopulation that lies within our boundary for 51~Eri and its median and 16th/84th percentiles correspond to 0.9$^{+0.3}_{-0.1}$ days. We show the y-axis at a limit of $\sim$5 days for aesthetic purposes but a few long-period outliers reside above this limit.  \label{fig:coreRot}} 
\end{figure*}

\citet{li20} used period-spacings for well-resolved $\gamma$~Dor pulsations to infer their core rotation periods. They also derived independent surface rotation periods for 58 of the stars based on their surface rotational modulation signal that was well-separated from pulsation modes to prevent mistaking the pulsations as rotation. They found that both rotation periods were consistent within 5\% for the entire subsample, suggesting that the core rotation is a reliable predictor of the surface rotation for these stars. Assuming that 51~Eri also rotates rigidly, a core rotation constraint can thus serve as a reliable tracer of the true surface stellar rotation period.

Figure \ref{fig:coreRot} shows an empirical relationship between the mean $\ell =1$ frequency and the core rotation periods for the population of $\gamma$~Dor stars from \citet{li20}. We estimated a mean $\ell =1$ frequency of 1.57$\pm$0.09 cycles/day for 51~Eri using the mean of the four significant $\ell =1$ frequencies from Table \ref{tab:freqs} and using the error of the mean as the uncertainty. We then used this uncertainty as a boundary to define a subpopulation sample of core rotation periods (N=81) that correspond to the \citet{li20} \textit{Kepler} $\gamma$~Dor stars whose mean $\ell =1$ frequency is consistent with that of 51~Eri (Figure \ref{fig:coreRot}). A histogram of these samples is shown as an inset in Figure \ref{fig:coreRot}. The median and 16th/84th percentiles of these samples correspond to a core rotation period of  0.9$^{+0.3}_{-0.1}$ days, which indicates that the true surface stellar rotation period could lie in this range. 

Until the core rotation for 51~Eri is directly measured, this histogram can be interpreted as a Bayesian prior for the expected rotation period of 51~Eri given our estimate of its mean $\ell =1$ frequency and given the empirical trends of the \textit{Kepler} $\gamma$~Dor population. We estimated the spin-axis inclination for 51~Eri using this 0.9 day stellar rotation period ($P_{\rm rot}$). We follow \citet{Maire+2019} in adopting a projected rotational velocity ($v$sin$i$) of 83 km/s, which is based on the spectroscopic $v$sin$i$ measurements from \citet{Royer+2007} and \citet{Luck2017}. We also used a radius of 1.67 \Rsun\ from combining the stellar angular diameter measurement of 51~Eri \citep[0.518 mas,][]{Simon+Schaefer2011} with the inverse \textit{Gaia} EDR3 parallax as the assumed distance. This results in an estimate of $\sim$62$^{\circ}$\footnote{In a similar fashion, if we instead assume an edge-on stellar inclination of $i = 90^{\circ}$ for 51~Eri \citep[which is similar to the orbital inclination of GJ~3305~AB,][]{Montet+2015} then this implies a maximum stellar rotation period estimate of 1.02 days. Truncating the core rotation period samples representative of 51~Eri (Figure \ref{fig:coreRot}) to only those that are also consistent with this maximum rotation period reduces the sample from N=81 to N=55 samples. The median and 16th/84th percentiles corresponding to this further reduced sample are 0.86$^{+0.09}_{-0.04}$ days.}. We did not propagate the statistical uncertainties of the input parameters for our estimation \citep{Masuda+Winn2020} nor included potential systematic uncertainty in the input $v$sin$i$ value (which may be influenced by the $\gamma$~Dor pulsations). In addition, the input $P_{\rm rot}$ is only an informed prediction. Therefore, we caution that our estimate of the stellar spin-axis inclination should not be treated as a robust  measurement. 

The core rotation period for 51~Eri may be measurable with future \textit{TESS} observations that extend the continuous time baseline to better resolve the $\gamma$~Dor pulsations. While such measurements typically require a long time baseline (e.g., we estimate $\sim$11 \textit{TESS} sectors are required to formally resolve a period spacing of $\sim$600~s, which is a typical spacing for stars with a similar mean $\ell =1$ frequency to 51~Eri), \citet{VanReeth+2022} recently demonstrated that two continuous sectors of \textit{TESS} photometry were required to measure the core rotation for the $\gamma$~Dor star HD~112429. This was largely a consequence of ideal period spacing and pulsation properties, and a longer time baseline would still be needed to improve precision and explore other interior properties. Nonetheless, this demonstrates that the prospect of a future core rotation constraint for 51 Eri may not be far-fetched.

\section{Conclusion \& Future Directions}\label{Sec:Conclusion}
We analyzed \textit{TESS} photometry to classify 51~Eri as a $\gamma$\,Dor pulsator. This star now joins HR~8799 as a directly-imaged exoplanet host star that is also a $\gamma$\,Dor variable. We noted that the previously quoted stellar rotation period of 0.65 days is most likely explained as pulsation modes, thereby making the stellar rotation period of 51~Eri presently undetermined. In light of this, the coplanarity of 51~Eri~b's orbit with the equitorial plane of its host star remains unknown. However, we are able to use our results to estimate a plausible rotation period of 0.9$^{+0.3}_{-0.1}$~days. We also found no significant evidence for transiting companions in our residual \textit{TESS} light curve. 

The detection of $\gamma$\,Dor pulsations makes 51\,Eri the only directly-imaged exoplanet host star thus far for which an \emph{unambiguous} asteroseismic age might be feasible. Additional \textit{TESS} data and multi-color ground-based photometry will be required to confirm the mode identification presented here and facilitate pulsation mode modeling to determine an asteroseismic age. An asteroseismic age would have implications for the $\beta$PMG in addition to all the constituents of the 51~Eri system. 

Additional \textit{TESS} data may also enable a measurement of the core rotation rate, which in turn will allow constraints on the surface rotation rate enabling a reevaluation of the stellar spin-axis inclination. Together with continued astrometric monitoring to reduce the uncertainty in the orbital inclination, this would thus also help exclude or make plausible a scenario where the orbit of 51~Eri~b is coplanar with its host star. Either case is a key detail in investigating the dynamical history of this system.

\textit{Acknowledgements}. We thank Travis Berger for crossmatching the \textit{Kepler} Input Catalog with \textit{Gaia} EDR3.

This material is based upon work supported by the National Science Foundation Graduate Research Fellowship Program under Grant No. 1842402. D.H. acknowledges support from the Alfred P. Sloan Foundation and the National Aeronautics and Space Administration (80NSSC21K0784). This research was funded in part by the Gordon and Betty Moore Foundation through Grant GBMF8550 to M. Liu. T.R.B. acknowledges support from the Australian Research Council (DP210103119).

This work has benefitted from The UltracoolSheet \citep{UCSZenodo}, maintained by Will Best, Trent Dupuy, Michael Liu, Rob Siverd, and Zhoujian Zhang, and developed from compilations by \citet{Dupuy+Liu2012}, \citet{Dupuy+Kraus2013}, \citet{Liu+2016}, \citet{Best+2018}, and \citet{Best+2021}. 

This paper includes data collected by the TESS mission, which are publicly available from the Mikulski Archive for Space Telescopes (MAST). Funding for the TESS mission is provided by the NASA's Science Mission Directorate.

This work has made use of data from the European Space Agency (ESA) mission {\it Gaia} (\url{https://www.cosmos.esa.int/gaia}), processed by the {\it Gaia} Data Processing and Analysis Consortium (DPAC, \url{https://www.cosmos.esa.int/web/gaia/dpac/consortium}). Funding for the DPAC has been provided by national institutions, in particular the institutions participating in the {\it Gaia} Multilateral Agreement. 

This research has made use of the SIMBAD database, operated at CDS, Strasbourg, France. 

This research has made use of NASA's Astrophysics Data System Bibliographic Services.

This research has made use of the VizieR catalogue access tool, CDS, Strasbourg, France (DOI : 10.26093/cds/vizier). The original description of the VizieR service was published in \citealt{Ochsenbein+2000}.
\facilities{TESS, Gaia}
\software{\texttt{lightkurve} \citep{LightkurveCollaboration+2018}, \texttt{SigSpec} \citep{SigSpec}, \texttt{matplotlib} \citep{matplotlib},  \texttt{pandas} \citep{pandas}, \texttt{astropy} \citep{astropy,astropy2018}, \texttt{numpy} \citep{Harris+2020}, \texttt{scipy} \citep{scipy}, \texttt{astroquery} \citep{astroquery}}

\bibliographystyle{aasjournal}
\bibliography{51Eri}

\end{document}